\documentclass[iop,apj,numberedappendix,twocolappendix,revtex4,letterpaper]{emulateapj}

\usepackage[pdfa,breaklinks,colorlinks,urlcolor=blue,citecolor=blue,linkcolor=blue,letterpaper]{hyperref}
\usepackage{natbib}
\usepackage{apjfonts}
\usepackage{amsmath}
\usepackage{graphicx}
\usepackage{bm}
\usepackage{paralist}
\usepackage[capitalize,noabbrev]{cleveref}
\usepackage{color}

\graphicspath{{figures/}}

\makeatletter
\usepackage{etoolbox}
\patchcmd\H@refstepcounter{\protected@edef}{\protected@xdef}{}{}
\makeatother
\usepackage{etoolbox}
\makeatletter
\patchcmd{\NAT@citex}
  {\@citea\NAT@hyper@{%
     \NAT@nmfmt{\NAT@nm}%
     \hyper@natlinkbreak{\NAT@aysep\NAT@spacechar}{\@citeb\@extra@b@citeb}%
     \NAT@date}}
  {\@citea\NAT@nmfmt{\NAT@nm}%
   \NAT@aysep\NAT@spacechar\NAT@hyper@{\NAT@date}}{}{}
\patchcmd{\NAT@citex}
  {\@citea\NAT@hyper@{%
     \NAT@nmfmt{\NAT@nm}%
     \hyper@natlinkbreak{\NAT@spacechar\NAT@@open\if*#1*\else#1\NAT@spacechar\fi}%
       {\@citeb\@extra@b@citeb}%
     \NAT@date}}
  {\@citea\NAT@nmfmt{\NAT@nm}%
   \NAT@spacechar\NAT@@open\if*#1*\else#1\NAT@spacechar\fi\NAT@hyper@{\NAT@date}}
  {}{}
\makeatother

\usepackage{xspace}

\newcommand{\mean}[1]{\ensuremath{\langle#1\rangle}}

\newcommand{\sub}[1]{\ensuremath{{_{\rm{#1}}}}}

\newcommand{\cps}{\,s$^{-1}$\xspace}

\newcommand{\xmm}{{\em XMM-Newton}\xspace}

\newcommand{\gammaray}{$\gamma$-ray\xspace}
\newcommand{\grays}{$\gamma$-rays\xspace}

\newcommand{\xray}{X-ray\xspace}
\newcommand{\xrays}{X-rays\xspace}

\newcommand{\sn}{$\rm{S/N}$\xspace}

\newcommand{\dof}{{dof\/}\xspace}

\newcommand{\fft}{{FFT\/}\xspace}
\newcommand{\ifs}{{IFS\/}\xspace}

\newcommand{\rk}{$\mathcal{R}^2_k$\xspace}
\newcommand{\z}{$\mathcal{Z}^2$\xspace}

\submitted{Received 2015 July 5; accepted 2016 February 11; published 2016 April 27}
\journalinfo{The Astrophysical Journal, 822:14 (11pp), 2016 April 27}

\begin{document}

\title{On More Sensitive Periodogram Statistics}
\shorttitle{Sensitive Periodogram Statistics}

\author{G.\ B\'elanger}
\affil{European Space Astronomy Centre (ESA/ESAC), Science Operations Department, Villanueva de la Ca\~nada (Madrid), Spain; \href{mailto:gbelanger@sciops.esa.int}{gbelanger@sciops.esa.int}}
\shortauthors{B\'elanger}

\begin{abstract}
Period searches in event data have traditionally used the Rayleigh statistic, $R^2$. For \xray pulsars, the standard has been the $Z^2$ statistic, which sums over more than one harmonic. For \grays, the $H$-test, which optimizes the number of harmonics to sum, is often used. These periodograms all suffer from the same problem, namely artifacts caused by correlations in the Fourier components that arise from testing frequencies with a non-integer number of cycles. This article addresses this problem. The modified Rayleigh statistic is discussed, its generalization to any harmonic, \rk, is formulated, and from the latter, the modified $Z^2$ statistic, \z, is constructed. Versions of these statistics for binned data and point measurements are derived, and it is shown that the variance in the uncertainties can have an important influence on the periodogram. It is shown how to combine the information about the signal frequency from the different harmonics to estimate its value with maximum accuracy. The methods are applied to an \xmm observation of the Crab pulsar for which a decomposition of the pulse profile is presented, and shows that most of the power is in the second, third, and fifth harmonics. Statistical detection power of the \rk statistic is superior to the FFT and equivalent to the Lomb--Scargle (LS). Response to gaps in the data is assessed, and it is shown that the LS does not protect against the distortions they cause. The main conclusion of this work is that the classical $R^2$ and $Z^2$ should be replaced by \rk and \z in all applications with event data, and the LS should be replaced by the \rk when the uncertainty varies from one point measurement to another.
\end{abstract}


\fontdimen2\font=0.65ex
\fontdimen3\font=0.65ex

\section{Introduction}

The power spectrum refers to the power spectral density distribution of a physical process, whereas the periodogram refers to an estimate of the power spectrum. The most common choice of a periodogram statistic is the Discrete Fourier Transform, and it is generally used in the form of a computationally fast algorithm called Fast Fourier Transform \citep[\fft; see][]{2002nrc..book.....P} that can only be applied to grouped data.

More sensitive periodogram statistics include the Rayleigh or $R^2$-test \citep{1983ApJ...272..256L}, the $Z^2$-test \citep{1983A&A...128..245B}, and the $H$-test \citep[][which automatically picks the optimal number of harmonics from which to compute $Z^2$]{1989A&A...221..180D}. Two important features that make these tests more powerful than the standard \fft periodogram are:
\begin{inparaenum}[(1)]
\item they can be applied directly to event arrival times, and thus access \emph{all} variability timescales present in the data, and 
\item they impose no constraints on the frequencies that can be tested, and are thus said to oversample the periodogram by testing timescales other than those corresponding to independent frequencies with an integer number of cycles.
\end{inparaenum}

However, oversampling without taking into account the fact that the variables computed to estimate the power at each frequency are correlated within each independent Fourier spacing (\ifs) leads to frequency-dependent artifacts that distort the periodogram and in some cases can be mistaken for, and interpreted as, the signature of a coherent periodic modulation. Each of the above-mentioned statistics---the $R^2$, $Z^2$ and $H$-test---suffers from this.

Although it is powerful---the most powerful according to \citet{1983ApJ...272..256L}---in detecting sinusoidal modulations in event data, the Rayleigh statistic achieves this by estimating the power using the fundamental harmonic only. This advantage in regards to strictly sinusoidal signals is a limitation when trying to detect, identify, or study non-sinusoidal pulse shapes. The $Z^2$ statistic was devised for this purpose as a generalization of the $R^2$ statistic that combines the power estimates from an arbitrary number of harmonics. Even if it is generally true, although not always the case, that the fundamental harmonic carries the bulk of the power, being able to access the additional power contained in higher harmonics confers the $Z^2$ statistic an important advantage over the $R^2$ statistic, and explains why it is the statistic of choice for event data where pulses are peaked or irregular in shape, as is often the case in pulsars.

A powerful and reliable periodogram statistic must fulfill three conditions: it must \begin{inparaenum}[(1)]
\item be able to use event arrival times in order to access all variability timescales, 
\item allow for oversampling in order to explore frequency space without restrictions, and
\item take into account the oscillation in the mean, variance, and covariance of the Fourier components as a function of frequency.\footnote{The expression "account for" and not "correct for" is used because the analytically predictable behavior of the oscillation in the value of the expected means, variances, and covariance is incorporated into the calculation. No correction is applied to the computed value of the statistic.}
\end{inparaenum} These criteria are met by the little known \emph{modified} Rayleigh statistic discussed in \cref{s:moreSensitiveStatistics}.

In light of these considerations, we introduce two new periodogram statistics: the generalized ($k$th order) modified Rayleigh statistic, labeled \rk; and the modified $Z^2$ statistic, labeled \z. Because these benefit from all the features of their predecessors but do not suffer from the artifacts caused by unaccounted for correlations in the trigonometric moments, it is probably most sensible to always use the \rk and \z instead of the $R^2$ and $Z^2$.

Just as the use of the $Z^2$ statistic can (and did) replace that of the $R^2$ in most applications, the new \z statistic should now be used in its stead in all event data applications, whether one is using solely the fundamental, reducing $Z^2$ to $R^2$ (as in \cite{2010MNRAS.401.1564R} who nevertheless cite \cite{1983A&A...128..245B} and not \cite{1983ApJ...272..256L}); using the first two (i.e., $\mathcal{Z}^2_2$ as originally suggested by \cite{1983A&A...128..245B}, and often used implicitly without actually specifying how many harmonics are used as in \cite{2012A&A...545A..83B}); or using several additional higher harmonics ($\mathcal{Z}^2_{10}$, for example, as was suggested by \cite{1986A&A...170..187D} and subsequently often used in period searches \cite[e.g.,][]{2009MNRAS.393..527D}). The \rk statistic that evaluates the contribution of the $k$th harmonic by computing the periodogram for that component is ideally suited for detailed investigations of non-sinusoidal pulse profiles in which the relative contributions of different harmonics to a complex profile are of interest.

This paper begins with a brief presentation of the standard power spectral estimation by \fft in which general notions relevant to spectral estimation are introduced (\cref{s:fft}). The classical $R^2$ and $Z^2$ periodograms and their limitations are then presented (\cref{s:r2AndZ2}) before turning to the consideration of the new \rk and \z statistics (\cref{s:moreSensitiveStatistics}), which are applied to an \xray observation of the Crab nebula, whose pulsar emission is characterized by a double, narrow-peaked, and thus highly non-sinusoidal pulse profile (\cref{s:crab}). The paper ends with some additional statistical considerations  (\cref{s:additionalConsiderations}) and a short conclusion (\cref{s:conclusion}). The derivation and examination of the \rk statistic are presented in the \hyperref[a:modRay]{Appendix}.

\section{Power Spectral Estimation by FFT}
\label{s:fft}

The \fft is performed on $n$ complex numbers, usually a power of 2, represented as an array of length $2n$ (each complex has a real and an imaginary part), and the operation yields $n$ complex numbers.  In the case of a light curve, the count rates per bin are real numbers, and therefore the imaginary parts are all zero.  The Fourier transform, $H_j$\,$\equiv$\,$\sum^{n-1}_{k=0} h_k e^{2\pi ikj/n}$, is periodic in $j$, with a period $n$, and symmetric about $n/2$.  It is computed for $(n+1)$ frequencies ranging from $-f_c$ to $f_c$, in steps of $\delta\/f$\,=\,$1/T$\,=\,$1/n\delta\/t$.  Here $f_c$ is the critical (highest) or Nyquist frequency defined as $f_c$\,$\equiv$\,$1/2\delta\/t$\,=\,$n/2T$, and $\delta\/t$ is the bin time of the input data.  Letting $j$ vary between 0 and $(n-1)$, we find that $H_j$\,=\,$H_{n-j}$.  Therefore, the transform yields $n/2$ distinct and meaningful complex numbers: the value of $j$\,=\,0 corresponds to a frequency of zero and equals zero when the mean is subtracted from the data prior to applying the transform; the value of $j$\,=\,$n/2$ corresponds to both $f_c$ and $-f_c$; and the values of 1\,$\leq$\,$j$\,$<$\,$n/2$ correspond to the positive frequencies.

The periodogram is constructed from the output of the transform by squaring the norm of each complex number and then applying a normalization.\footnote{Even though the input data (a time series of intensities) are real with all imaginary parts equal to 0, the output of the operation is complex, and so the norm is the complex modulus.}  Common choices include the Leahy normalization \citep{1983ApJ...266..160L} that places the white noise level at 2, and the fractional variance normalization \citep{1990A&A...227L..33B,1991ApJ...383..784M}, where the integral of the periodogram between two frequencies yields the square of the fractional {\small RMS\/} contribution in that range. The \fft is fast and ideal in many applications.

The duration of the observation determines both the minimum and the step between independent frequencies. The number of \ifs in the frequency range [$\nu\sub{min}:\nu\sub{max}$] is given by  $T\sub{obs}(\nu\sub{max} - \nu\sub{min})$ or equivalently by $T\sub{obs}(1/P\sub{min} - 1/P\sub{max})$. This defines all independent frequencies up to the critical (maximum) derived from the sampling interval (bin time). A shorter bin time translates into a higher critical frequency, and thereby extends the sampling to higher frequencies; sampling of low frequencies remains unchanged. In this form, there is no sampling between independent frequencies, and this restricts the ability to detect weak signals.

\section{The Classical $R^2$ and $Z^2$ Statistics}
\label{s:r2AndZ2}

One great strength of the Rayleigh and $Z^2$ statistics is that they are computed directly from the time of arrival of events, which allows the estimation of the power spectrum using the distribution in time of these events exactly as detected, no matter how many or how few, without grouping, and without restrictions on which frequencies can be tested. This yields a more sensitive way to detect periodic signals, particularly weak signals, especially if the periodicity happens to be exactly between two independent frequencies.

\begin{figure*}[!ht]
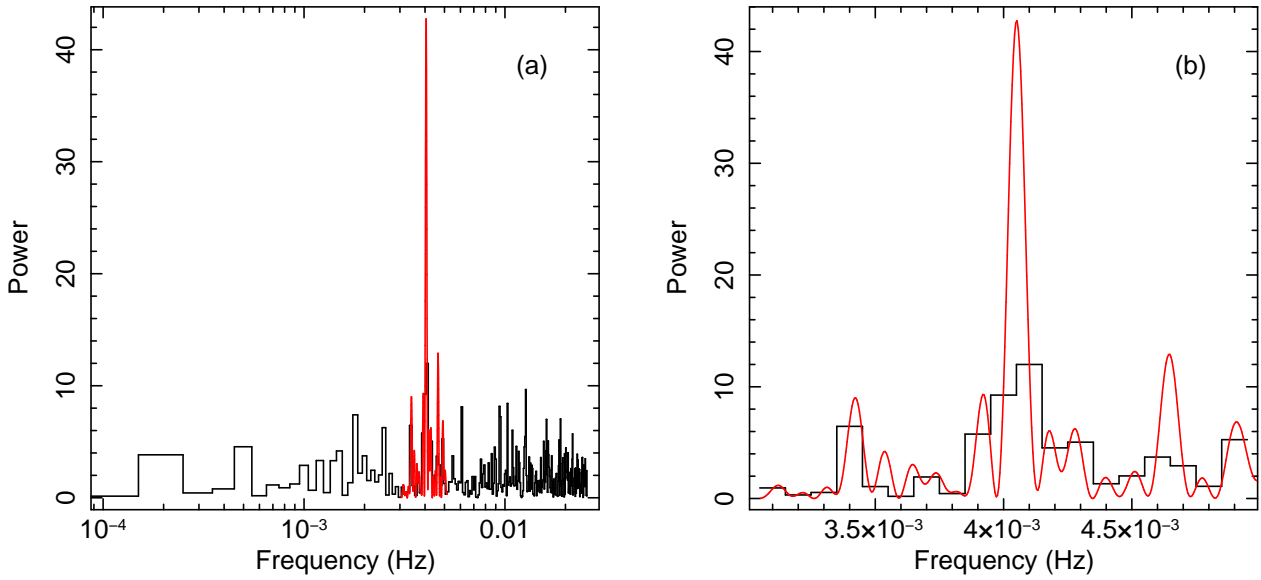

\epsscale{1.0}
\includegraphics[height=0.45\textwidth, angle=-90]{fftAndR2_narrowRange.ps} \quad
\includegraphics[height=0.45\textwidth, angle=-90]{fftAndR2_zoom.ps}
\caption{\footnotesize Using the $R^2$ statistic to search for weak periodic signals. The simulated data are of white noise (duration $T$\,=\,10\,ks, mean rate $\mu$\,=\,0.5\cps), with a 10\% pulsed fraction (1 of 10 events is modulated; \cref{eq:snr}) for a sinusoid at 0.00405\;Hz ($\approx$\,247\,s). The FFT periodogram in black is computed on 512 time bins ($\delta t$=19.531\,s) and thus 256 real frequencies with $\nu\sub{min}$\,=\,$\delta\/f$\,=\,$10^{-4}$\,Hz and $\nu\sub{max}$\,=\,0.0256\,Hz. The Rayleigh periodogram in red is around $\pm$1 IFS of the peak (0.003--0.005 Hz), with sampling of 21 frequencies per IFS. Panel (a) shows the full range, and panel (b) shows the range of the $R^2$ periodogram. This example with a period between two independent frequencies was picked to clearly illustrate the important difference in sensitivity that can be achieved in some cases.}
\label{f:fftAndR2}
\end{figure*}

The $R^2$ statistic uses only the fundamental harmonic, and is therefore most sensitive to sinusoidal signals, but performs quite well for any kind of periodic pulsations. The $Z^2$ extends the $R^2$ by summing over more than one harmonic, and is therefore more versatile at detecting non-sinusoidal pulse shapes. It also allows, for example, monitoring of the power in the second harmonic in order to study the time evolution of the signal when the power associated with the fundamental is known to fluctuate due to factors unrelated to the signal's periodic nature \citep[e.g.,][]{2006ApJ...653L.133B}.

\subsection{The R\textsuperscript{2} Statistic}
\label{s:r2}

From a data set comprising $N$ events, the Rayleigh power at a given frequency $\nu$ (or period $P$), is calculated by converting each arrival time, $t_i$, to a phase, $\phi_i$, given by the fractional part of $2\pi t_i \nu$ (or $2\pi t_i/P$), and computing
\begin{equation}
\label{eq:r2}
R^2 = \frac{2}{N} \left[ \left(\sum_{i=1}^{N} \cos\phi_i \right)^2 + \left(\sum_{i=1}^{N} \sin\phi_i \right)^2 \right].
\end{equation}

\cref{f:fftAndR2} shows the Rayleigh (red) and \fft (black) periodograms of a simulated white noise with a sinusoidal modulation, showing the full \fft periodogram (panel (a)) and a zoom that highlights the details of the Rayleigh power estimates around the peak (panel (b)). This particular example was picked to illustrate the sometimes remarkable difference in the sensitivity of these two statistics in regards to a weak modulation exactly between two independent frequencies. In the \fft (power of 11.7 at 0.0040\;Hz and 9.3 at 0.0041\;Hz), for which it is worth noting that it would most probably not have been identified as unusual given the presence of comparable fluctuations in the power throughout the periodogram, compared to the $R^2$ in which the signal stands out without a doubt, either about its presence (power of 42.8), its peak location (at 0.004052\,Hz), or its low probability of having arisen from a random fluctuation ($10^{-10}$). 

\subsection{The Z\textsuperscript{2} Statistic}
\label{s:z2}

Multiplying the argument of the sine and cosine functions by an integer greater than one when computing $R^2$ will probe the data for the contribution from higher order harmonics.  Summing over more than one of these yields the $Z^2$ statistic, usually labeled $Z^2_m$ to indicate the number of harmonics $m$ included in the sum:
\begin{equation}
\label{eq:z2}
   Z^2_m = \frac{2}{N} \sum_{\rm k=1}^{m}
   \left[
	\left(\sum_{i=1}^{N}\cos(k\phi_i)\right)^2 + 
	\left(\sum_{i=1}^{N}\sin(k\phi_i)\right)^2
   \right].
\end{equation}

\noindent Here also, $N$ is the number of events, and $\phi_i$ is the phase of the event with arrival time $t_i$, but in addition, we have the integer variables $k$---as the index of the harmonic, and $m$---as the total number of harmonics included in the power estimate.

While the main distinctions between $Z^2$ and $R^2$ are the former's sensitivity to non-sinusoidal signals with arbitrary pulse shapes and its ability to look at harmonics beyond the fundamental, another distinction arises from the fact that summing the contribution to the power estimates from more than one harmonic equates to summing as many periodograms as there are harmonics in the sum. Therefore, the sampling distribution of the power estimates---and hence the statistics---depends on this number. Each periodogram has powers that are distributed as a $\chi^2_2$ variable for white noise. This implies that (for white noise) summing two, three, or four harmonics will yield powers distributed as a $\chi^2_4$, $\chi^2_6$, or $\chi^2_8$ variable. This is a distinction between $R^2$ and $Z^2_m$ that is particularly important when estimating the probability or evaluating the likelihood of a particular value of power in the periodogram. Summing more than one harmonic also decreases the variance because random statistical scatter is averaged over more than one periodogram.

\subsection{Artifacts in the R\textsuperscript{2} and Z\textsuperscript{2} Periodograms}
\label{s:artifacts}

Unfortunately, as sensitive as the $R^2$ statistic may be to weak sinusoidal signals, and as sensitive as the $Z^2$ statistic may be to non-sinusoidal pulse profiles, both suffer in exactly the same way from oversampling artifacts caused by correlations within each \ifs throughout the periodogram, but that are most noticeable at lower frequencies (\cref{f:oversamplingartifacts}, panel (a)).

\begin{figure*}[!ht]
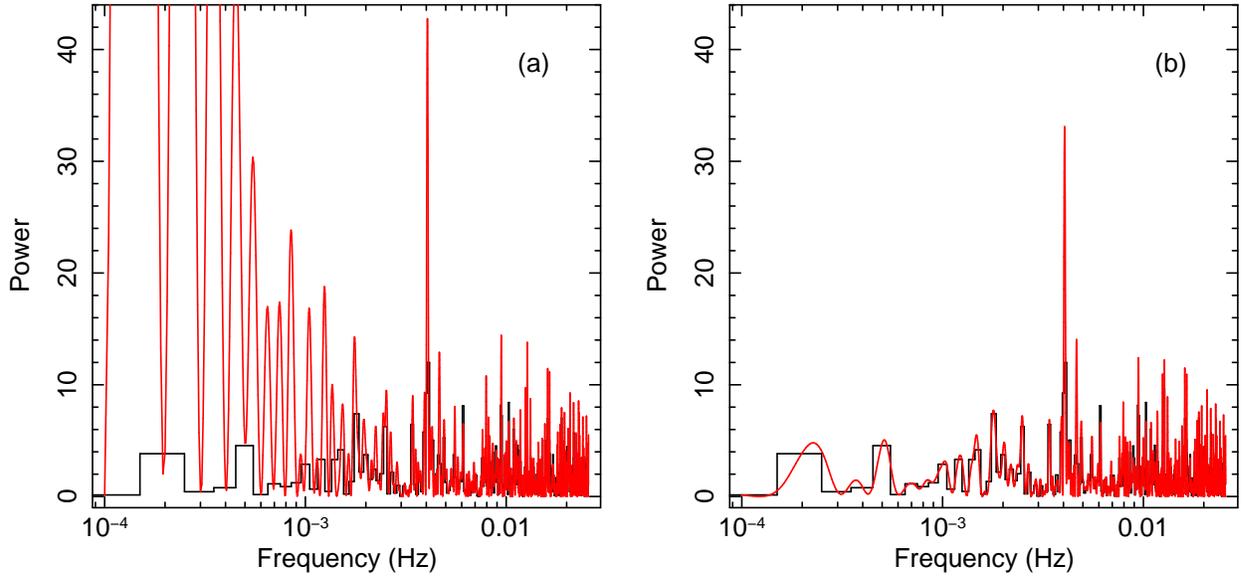

\epsscale{1.0}
\centering
\includegraphics[height=0.45\textwidth, angle=-90]{fftAndR2_fullRange.ps}
\includegraphics[height=0.45\textwidth, angle=-90]{fftAndModR2_fullRange.ps}
\caption{\footnotesize Artifacts in the $R^2$ periodogram (in red, computed using the same data as in \cref{f:fftAndR2}) are shown in panel (a) on a truncated linear scale, visibly growing in a power-law fashion toward lower frequencies, with $R^2$ estimates between independent frequencies deviating noticeably from the FFT estimates below $\approx 3\times10^{-3}$\,Hz. In panel (b) we see the $\mathcal{R}^2_1$ statistic applied to the same data, with the periodic signal clearly detected at the same frequency (0.004052 Hz) but at a somewhat lower power from the more accurate calculation (33.1 instead of 42.8, and a probability of $10^{-8}$ instead of $10^{-10}$ of arising from a noise fluctuation).}
  	\label{f:oversamplingartifacts}
\end{figure*}

For independent frequencies (those with an integer number of cycles), the integral of the sine and cosine components is always zero. For all other frequencies, this is not the case, and the value of the integral oscillates between the independent frequencies.  This is similarly true for their variances (assumed to be equal to one-half) and covariance (assumed to be zero), all of which also oscillate. As expected, then, powers in the \fft and $R^2$ periodograms are equal or nearly so at independent frequencies, but can vary wildly in between.

\section{More Sensitive Periodogram Statistics}
\label{s:moreSensitiveStatistics}

Fortunately, because the integral of the Fourier components, as well as their variances and covariance, can be computed exactly, it is possible to account for the fluctuations in their value and thereby eliminate the artifacts that they produce. A modification to the $R^2$ statistic that takes into account the expected means and variances (but not the covariance) to standardize the Fourier moments was used (but not emphasized) by \citet[][Equations (8)--(10)]{1994ApJ...436..239D} in a search for pulsed emission from \gammaray pulsars. A full correction that accounts for expected means, variances, and covariance was presented independently by \citet[][Section 4.1]{1996APh.....4..235O}. The latter's modified Rayleigh statistic, which we label $\mathcal{R}^2$, is much more sensitive than the classical version because it does not suffer from artifacts as does the classical $R^2$ statistic. It was formulated for the fundamental harmonic, and this limits its applicability to detailed studies with high-quality data.

\subsection{The new \rk and \z Statistics}
\label{s:mod-r2_k}

We define the generalization of the modified Rayleigh statistic for any harmonic as
\begin{equation}
\label{eq:r2k}
	\mathcal{R}^2_k = 
	\begin{pmatrix}
	  C_k - \mean{C_k} \\ 
	  S_k - \mean{S_k}
	\end{pmatrix}^{\rm T}
	\begin{pmatrix}
	   {\sigma^2}_{C_k} & \sigma_{C_k S_k} \\
	    \sigma_{C_k S_k}   & {\sigma^2}_{S_k}
	\end{pmatrix}^{-1}
	\begin{pmatrix}
	  C_k - \mean{C_k} \\ 
	  S_k - \mean{S_k}
	\end{pmatrix}
\end{equation}
The dependency on the harmonic is carried by the variable $k$ in the argument of the sine and cosine functions to yield the following expressions for $C_k$ and $S_k$:
\begin{equation}
\label{eq:fourierComp2}
	C_k = \frac{1}{N} \sum_{i=1}^N \cos{k\phi_i}
	\hspace{4mm} {\rm and} \hspace{4mm}
	S_k = \frac{1}{N} \sum_{i=1}^N \sin{k\phi_i}.
\end{equation}
The other terms are defined as follows:
\begin{eqnarray}
\label{eq:expectedC2}
\mean{C_k}	& = & \frac{1}{k\omega\/T} \left[ \sin{k\omega\/t} \right]_{t_1}^{t_2}, \\
\mean{S_k}	& = & \frac{-1}{k\omega\/T} \left[ \cos{k\omega\/t} \right]_{t_1}^{t_2}, \\
\sigma^2_{C_k} & = & \frac{1}{2N} \left( 1 + \frac{1}{k\omega\/T} \left[ \sin{k\omega\/t}\cos{k\omega\/t} \right]_{t_1}^{t_2} \right) - \mean{C_k}^2, \\
\sigma^2_{S_k} & = & \frac{1}{2N} \left( 1 - \frac{1}{k\omega\/T} \left[ \sin{k\omega\/t}\cos{k\omega\/t} \right]_{t_1}^{t_2} \right) - \mean{S_k}^2, \\
\sigma_{C_k S_k} & = & \frac{1}{2k\omega\/TN} \left[ \sin^2k\omega\/t \right]_{t_1}^{t_2} - \mean{C_k} \mean{S_k}.
\label{eq:expectedFourierComp2}
\end{eqnarray}
The terms $\mean{C_k}$ and $\mean{S_k}$ are the expectation values, $\sigma^2_{C_k}$ and $\sigma^2_{S_k}$ are the variances, and $\sigma_{C_k S_k}$ is the covariance of $C_k$ and $S_k$. (See the  \hyperref[a:modRay]{Appendix} for the details of the derivation.)

\cref{f:oversamplingartifacts} panel (a) shows the $R^2$ periodogram, and panel (b) shows the modified Rayleigh statistic and demonstrates the advantage it has over the standard \fft periodogram for detecting weak signals peaking between independent frequencies without the severely limiting disadvantages of the classical Rayleigh statistic. \rk is identically as sensitive as $\mathcal{R}^2$ for the fundamental harmonic (by mathematical definition), but it is, in addition, equally sensitive for any other harmonic.

\begin{figure*}[!ht]
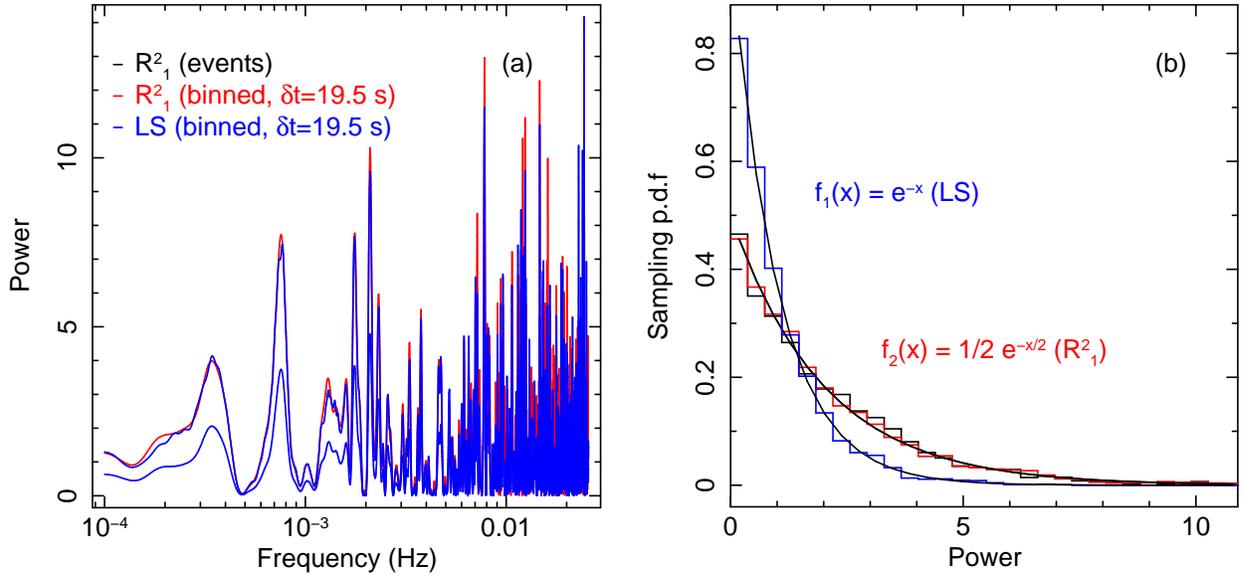

\epsscale{1.0}
\centering
\includegraphics[height=0.45\textwidth, angle=-90]{modR2ForEventsAndTS_withLomb.ps}
\includegraphics[height=0.45\textwidth, angle=-90]{modR2ForEventsAndTS_withLomb-histo.ps}
  \caption{\footnotesize Comparison in panel (a) of the \rk statistic for event data (\cref{eq:r2k}, in black) and binned data (\cref{eq:r2k1-lc}, in red), with the LS statistic (\cref{eq:lomb}, in blue), showing the corresponding sampling distributions in panel (b), all of which follow the expectation (regardless of the sampling factor).}
  \label{f:modR2ForEventsAndTS_withLomb}
\end{figure*}

Because the $Z^2$ periodogram is a sum of $R^2_k$ components, it is therefore natural to use the much more sensitive \rk as the kernel for a new, \emph{modified} $Z^2$ statistic, \z, defined as
\begin{equation}
\label{eq:mod-z2}
   \mathcal{Z}^2 = \sum\mathcal{R}^2_k.
\end{equation}
Summing as many harmonics as deemed necessary depending on the profile (or expected profile) ensures that all the power present in the selected harmonics of the periodic modulation is included and yields the highest possible peak in the final \z periodogram.\footnote{Note that the selection of harmonics summed is not restricted in any way: it can be from 1 to $m$, as it has been used in the classical $Z^2_m$ statistic; it can be sequential from $i$ to $m$, in the case where the fundamental is glitchy, for example; or it can be an arbitrary set of harmonics selected according to some other criteria (2, 3, and 5, for example). This is why we do not specify the summation indices in \cref{eq:mod-z2,eq:fw,eq:sigf}.} But because different harmonics will be distributed differently around the peak, they will in general also reach their maxima at slightly different frequencies. Therefore, to derive a single best estimate of the signal's frequency, we need a way to combine the information carried by each harmonic that maximizes accuracy. For this purpose we construct a weighted mean statistic given by
\begin{equation}
\label{eq:fw}
   f_w = \frac{\sum w_k f_k}{\sum w_k},
\end{equation}
where the weight of each term, $w_k$, is the ratio of the peak power ($\mathcal{R}^2_k({\rm peak\/})$) to the square of the peak's half width at half maximum ($\sigma^2_k$). The motivation is to give more weight to the value of the peak frequency derived from taller and narrower peaks. The $k$th harmonic is therefore weighted by
\begin{equation}
\label{eq:w}
   w_k = \frac{\mathcal{R}^2_k({\rm peak\/})}{\sigma^2_k}.
\end{equation}

The uncertainty on the resulting frequency is calculated using the peak widths of the contributing harmonics:\footnote{The precision with which each peak frequency is determined depends on the width of its peak and not on its height.}
\begin{equation}
\label{eq:sigf}
 	\sigma_{f_w} = \frac{1}{\sqrt{\sum \sigma^2_k}},
\end{equation}
where the sum is over the same harmonics as those used in \cref{eq:fw} to compute the weighted mean frequency. The use of these is illustrated in \cref{s:crab}.

The periodogram statistics \rk and \z of Equations (\ref{eq:r2k})--(\ref{eq:expectedFourierComp2}) and (\ref{eq:mod-z2})--(\ref{eq:sigf}) are formulated for event data and thus cannot be used on binned or point measurement data. We therefore introduce a suitable formulation for such cases.

\subsection{The \rk for Binned Data and Point Measurements}
\label{s:modRayForLC}

A great part of the strength of an event data periodogram statistic is that it is computed directly from the event arrival times. This gives us access to the highest frequencies that cannot be investigated when grouping the data on an even slightly longer timescale.  Another important feature of the \rk is that it allows for the testing of as many frequencies as desired without distorting the periodogram (i.e., without distorting the sampling distribution); this is a feature that is essential when searching for weak, low-frequency signals.  A version of the \rk and (by extension) \z statistics for binned data or point measurements of an intensity with an associated uncertainty (as in radio data, for example) extends the use of these sensitive periodograms to data other than lists of arrival times.

This is the \rk statistic for light curves:
\begin{equation}
	\mathcal{R}^2_k =
\frac{\left(\sum_{i=1}^{n} \rho_i \cos{k\phi_i} \right)^2}{\sum_{i=1}^{n} (\rho_i\cos{k\phi_i})^2}
+ \frac{\left(\sum_{i=1}^{n} \rho_i \sin{k\phi_i} \right)^2}{\sum_{i=1}^{n} (\rho_i \sin{k\phi_i})^2}
\label{eq:r2k-lc}
\end{equation}
The sum is performed over $n$---the number of bins in the time series, instead of on $N$---the number of events in the list. The phases, $\phi_i$, are calculated using the time of the measurement or the center of the bin, and the $\cos{k\phi_i}$ and $\sin{k\phi_i}$ terms are weighted by $\rho_i = r_i/\sigma^2_i$ defined as the mean-subtracted intensity, $r_i$, divided by the square of the uncertainty, $\sigma^2_i$, for each measurement.\footnote{The mean-subtracted intensity tends to a zero-centered normal variable. For this reason, the expectation of $\rho_i \cos{k\phi_i}$ and $\rho_i \sin{k\phi_i}$ is also zero, and thus there are no artifacts due to a non-zero expectation. Normalization by the denominator terms results in a sum of two squared standard normal variables that yields a $\chi^2_2$ distributed variable (for white noise).}

For the simplest case, $k=1$, we get the modified Rayleigh or $\mathcal{R}^2_1$ statistic for binned data and point measurements:
\begin{equation}
	\mathcal{R}^2_1 =
\frac{\left(\sum_{i=1}^{n} \rho_i \cos\phi_i \right)^2}{\sum_{i=1}^{n} (\rho_i\cos\phi_i)^2}
+ \frac{\left(\sum_{i=1}^{n} \rho_i \sin\phi_i \right)^2}{\sum_{i=1}^{n} (\rho_i \sin\phi_i)^2}
\label{eq:r2k1-lc}
\end{equation}
The $\mathcal{R}^2_1$ for light curves is reminiscent of the Lomb--Scargle (LS) statistic \citep{1982ApJ...263..835S}:
\begin{equation}
   \label{eq:lomb}
      \mathcal{P}(\omega) = \frac{1}{2\sigma^2}\!\!
	\left[\!
		\frac{\left(\sum r_i \cos\omega(t_i - \tau) \right)^2}{\sum \cos\omega(t_i - \tau)}
		+ \frac{\left(\sum r_i \sin\omega(t_i - \tau) \right)^2}{\sum \sin\omega(t_i - \tau)}
	\right]\qquad
\end{equation}
where, as above, all sums on $i$ are from 1 to $n$; $\sigma^2$ is the variance of the mean-subtracted rates; $r_i$ is the mean-subtracted count rate in bin $i$; $\omega=2\pi/P=2\pi\nu$ is the angular frequency; $t_i$ is the time at the bin center; and $\tau$ is defined according to $\tan(2\omega\tau) = \sum_{i=1}^{n} \sin{2\omega t_i} / \sum_{i=1}^{n} \cos{2\omega t_i}$ \citep[see][]{2002nrc..book.....P}.

As is shown in panel (a) of \cref{f:modR2ForEventsAndTS_withLomb}, the periodograms resulting from applying the $\mathcal{R}^2_1$ or LS statistics to white noise data are virtually identical in shape. The LS differs by a factor of two in normalization that translates into different sampling distributions shown in panel (b): $f(x)$\,=\,$\onehalf\,e^{-x/2}$ for the $\mathcal{R}^2_1$, and $f(x)$\,=\,$e^{-x}$ for the LS statistic.\footnote{Removing the factor of one-half in the normalization of the LS yields a sampling distribution identical to that of the $\mathcal{R}^2_1$ for gapless white noise. See also \citet{2013A&C.....1....5V} for a detailed discussion of the LS periodogram.} Another difference is that the popular and commonly used LS periodogram of \cref{eq:lomb} does not take into account uncertainties \cite[but see][]{1989ApJ...343..874S}.

It is important to highlight that the size of the uncertainties, no matter how large this may be, does not affect the shape of the periodogram. It is the \emph{variance} in the values of uncertainties---the magnitude of the variations between individual uncertainties in the set of measurements---that affects the periodogram. This is illustrated in the \hyperref[a:modRay]{Appendix}.

\section{Multi-harmonic Decomposition of the Crab Pulsar's X-ray Pulse Profile}
\label{s:crab}

The Crab pulsar, with its highly peaked and asymmetric pulse profile, is an excellent example for which it is not only useful but necessary to use several harmonics to characterize the pulse shape and accurately estimate the spin period. We use an \xmm observation to illustrate in practical terms the use of the statistics introduced above for timing studies of pulsars with non-sinusoidal profiles. The \rk statistic (Equations (\ref{eq:r2k}) or (\ref{eq:r2k-lc})) allows us to compute the periodogram for each harmonic individually and thus see each one's relative contribution to the pulse, and Equations (\ref{eq:fw})--(\ref{eq:sigf}) allow us to most accurately compute the pulse frequency and the uncertainty on its value by combining statistically the information from each of the individual harmonics.

\begin{figure}[!ht]
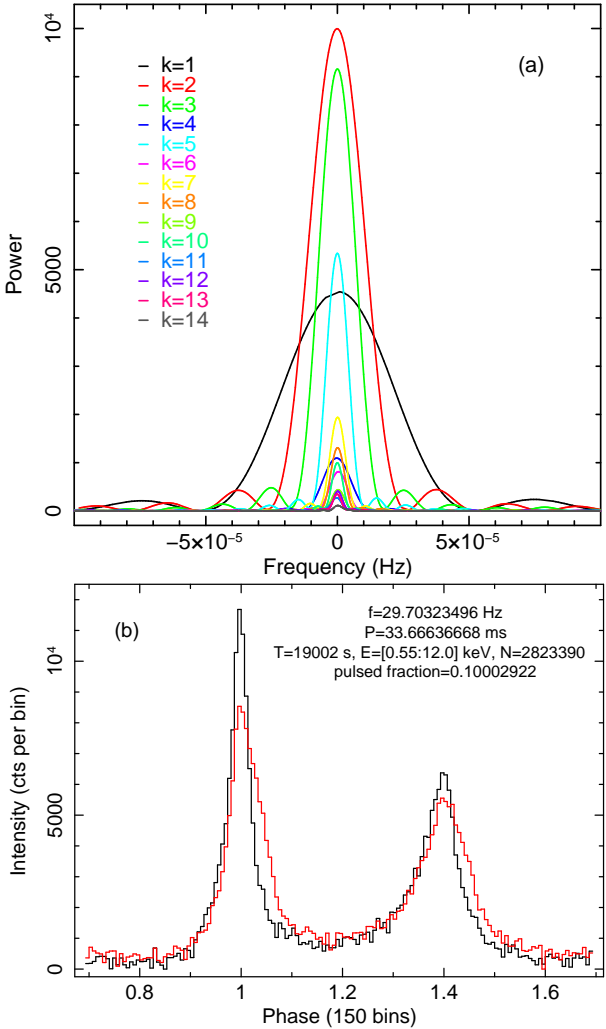

  \epsscale{1.0}
    \hspace{-2mm}\includegraphics[height=0.955\columnwidth, angle=-90]{mod_psd-crab-pn.ps}
    \includegraphics[height=0.935\columnwidth, angle=-90]{phasoBestAndRadio_0611181501.ps}
  \caption{\footnotesize Multi-harmonic periodogram of the Crab pulsar's X-ray emission centered on the pulse frequency (a) and phasograms resulting from folding the arrival times on our best estimate of the pulse frequency compared to the frequency derived from the radio ephemerides (b). The data is from an \xmm observation (ID 0611181501-003 on 2012 February 24--25) with an elapsed time of 19,002 s, using the Epic PN Timing/FastBurst data comprising 2823390 events in the range 0.55--12.0 keV (mean rate 148.6\cps).}
  \label{f:crab}
\end{figure}

\cref{f:crab} shows (in panel (a)) the multi-harmonic decomposition of the Crab pulsar's pulse up to $k = 14$ where we see that the highest peaks are those of the second, third, and fifth harmonics, and (in panel (b)) the profiles that result from folding the data at our best estimate of the \xray pulse frequency (in black) from the first five harmonics (details in \cref{t:peaks}), or folding at the frequency derived from the Jodrell Bank ephemerides for the midpoint of the observation (in red).

\begin{deluxetable}{cccc}
\tablecolumns{4} 
\tablewidth{0pc} 
\tablecaption{Characteristics of the First Five Harmonics} 
\tablehead{ 
\colhead{Harmonic} & \colhead{Peak Height} &  \colhead{Frequency} & \colhead{HWHM} \\
	(k)			&($\mathcal{R}^2$ Power)	& \colhead{(Hz)}	& \colhead{($10^{-6}$\,Hz)}}
\startdata
1 & 4537 & 29.70323424 & 23.81 \\
2 & 9988 & 29.70323534 & 12.19 \\
3 & 9155 & 29.70323486 & 7.764 \\
4 & 1097 & 29.70323461 & 5.847 \\
5 & 5342 & 29.70323499 & 4.689 \\
1--5$^{~\mathrm{a}}$ & 30119 & 29.70323496 & 3.165
\enddata 
\tablenotetext{a}{The peak height is the sum of powers; the frequency and uncertainty are those given by Equations (\ref{eq:fw})--(\ref{eq:sigf}).} 
\label{t:peaks}
\end{deluxetable} 

\section{Additional Statistical Considerations}
\label{s:additionalConsiderations}

\subsection{Signal Detection with the $\mathcal{R}^2$ Statistic}
\label{s:modRayPower}

It was shown in \cref{f:fftAndR2} that under certain conditions the power at a given frequency can be underestimated to the extent of not being identified as an interesting feature. The performance of a statistic  can be evaluated using simulations.\footnote{The accuracy depends only on the number of synthetic data sets.} We simulate data that contain a signal of specified signal-to-noise ratio (\sn; \cref{eq:snr}), and count how often a detection would be claimed given a particular threshold. The statistics are compared on an equal footing because the data are the same. We can estimate the probability of false negatives, $\beta$ (missing the signal that is there: a type II error), often of greater interest when searching for weak signals, as well as the probability of false positives, $\alpha$ (claiming a detection when there is no signal: a type I error). The fraction of false negatives is $1-\hat{\beta}$, and the fraction of false positives is $1-\hat{\alpha}$. The former quantifies the statistic's sensitivity (how often it detects a signal that \emph{is} in the data), and the latter quantifies its reliability (how often it detects a signal that \emph{is not} there).%
\footnote{The Neyman-Pearson system refers to $\alpha$ as the size, and $1-\beta$ as the power of the statistic or procedure. It is interesting to note that in testing hypotheses, the likelihood ratio of a measured value under one hypothesis versus the other is equal to $(1-\beta)/\alpha$. More explicitly, the statement `a test of $H_1$ versus $H_2$ having size $\alpha$ and power $1-\beta$ led to the acceptance of $H_1$' is evidence favoring $H_1$ by the factor $(1-\beta)/\alpha$, even if it does not necessarily lead to a proper evidential interpretation of the observed value \citep[see][p.\ 49]{Royall:1997vc}.}

If the \sn of the sinusoidal signal is high, any periodogram will detect the modulation. The considerations herein are therefore of practical relevance only for weak signals, particularly when they are at low frequencies for which the $\mathcal{R}^2_1$ is best suited. For event data the \sn is 
\begin{equation}
\label{eq:snr}
\rm{S}/\rm{N} = \frac{S}{\sqrt{S+B}} = \frac{\pi\/N}{\sqrt{N}} = \pi\sqrt{N}, \\
\end{equation}
where $S$ and $B$ respectively stand for the number of signal and background events, $N$ for their sum (the total number of events), and $\pi$ for the pulsed fraction. We consider sinusoids.\footnote{The quality of the data in this regime does not allow for a meaningful study of the pulse shape that can be severely distorted due to the low statistics. Therefore, as long as the pulse is relatively broad, the results are also applicable to non-sinusoidal pulses.} 

\begin{figure*}[!ht]
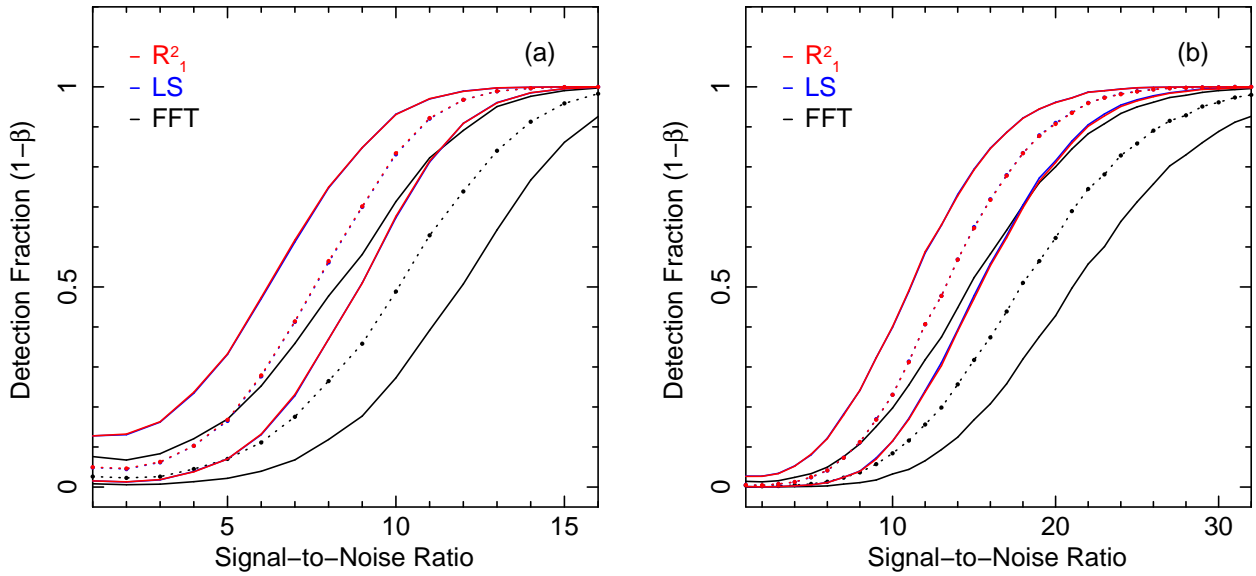

\epsscale{1.0}
\includegraphics[height=0.45\textwidth, angle=-90]{detectFracVsSNR_t_10000.0_mu_0.5_P_2857.0_alpha_0.0.ps}\quad
\includegraphics[height=0.45\textwidth, angle=-90]{detectFracVsSNR_t_10000.0_mu_0.5_P_2222.0_alpha_1.0.ps}
  \caption{\footnotesize Detection fraction as a function of \sn for a sinusoidal modulation shown in panel (a) for a period of 2857\,s (3.5 cycles in 10 ks) in a white noise background, and in panel (b) for a period of 2222\,s (4.5 cycles) in red noise with $\alpha=1$. The curves are derived from 10,000 runs for each \sn value (marked by dots on the dashed curves). Each run produces a number of background and signal events based on the pulsed fraction, $\pi$ (\cref{eq:snr}). The average background power is determined in the range $\pm 3$\,IFS, and the peak search is done within $\pm 1$\,IFS. The thresholds are defined as the likelihood ratios of the normal density at 2.5$\sigma$, 3.0$\sigma$, and 3.5 $\sigma$ with respect to its mode at zero: they are equal to $4.4\times10^{-2}$, $1.1\times10^{-2}$, and $2.2\times10^{-3}$, respectively. The detection fraction is the number of peaks (out of 10,000) with a likelihood ratio equal to or less than the threshold.}
  \label{f:detectionRateVsSNR}
\end{figure*}

\cref{f:detectionRateVsSNR} shows the results of simulations done to estimate the proportion of false negatives as a function of \sn for white and red noise in the limiting case of the lowest detectable frequency for which the period is right between two independent test frequencies. The comparison is between the \fft (black), the LS (blue), and the $\mathcal{R}^2_1$ periodograms (red), and the results are presented using likelihood intervals between 2.5$\sigma$ (upper) and 3.5$\sigma$ (lower curves), with the dashed lines at 3$\sigma$ (in the Gaussian sense).

The $\mathcal{R}^2_1$ and LS (scaled by a factor of 2) perform identically well, and significantly better than the \fft. The difference between having a white or red noise background is substantial: for example, a 97\% detection fraction at 3$\sigma$ is reached at an \sn of 11 for white noise, but at an \sn of 21 for red noise. Because we rely on the likelihood ratio to compare against the average background, larger fluctuations in the periodogram caused by the red noise affect detection sensitivity ($1-\hat{\beta}$) but not reliability ($1-\hat{\alpha}$). Calculated in the same way but using background events only, the false detection fraction is very low at around 1.7\%, and even more importantly, it is practically the same for white noise as it is for the $\alpha=1$ red noise.\footnote{In addition, calculating this fraction as a function of sampling factor for the LS and $\mathcal{R}^2_1$ shows that it remains constant, independent of the number of frequencies tested within an IFS. This implies, at least for $1-\hat{\alpha}$,  that the practice of "correcting for the number of trials" by dividing the probability associated with the power at the peak by the number of test frequencies \citep[e.g,][]{2008ApJ...688L..17M} is unnecessary (and incorrect because it is an over-correction) since likelihood ratios do not change with the number of trials or draws. A good practice is to evaluate $1-\hat{\alpha}$ and $1-\hat{\beta}$ for the statistic we are using and type of data we are working with in order to objectively report on the performance of the method.}

\subsection{Effects of Data Gaps}
\label{s:dataGaps}

Gaps in the data introduce structures in the periodogram. To constrain and quantify these effects cannot be done in a general way because they depend on the data and on the size and distribution of gaps within the data. Carrying out a study to investigate gap-induced modifications of the periodogram for any number of gap structures and data sets, real or simulated, could be useful and informative (maybe even essential) depending on the application. But in order to be of practical use, the study would have to be focused on the problem at hand, which would  define the parameter space to be explored based on the instrumental and observational features of the telescope, and the physical and statistical characteristics of the source.  Here we want to illustrate the effects of gaps for a few simple cases only to get an idea of their magnitude and of the performance of the LS statistic---specifically formulated to handle gaps---compared to the $\mathcal{R}^2_1$.

\begin{figure*}[ht]
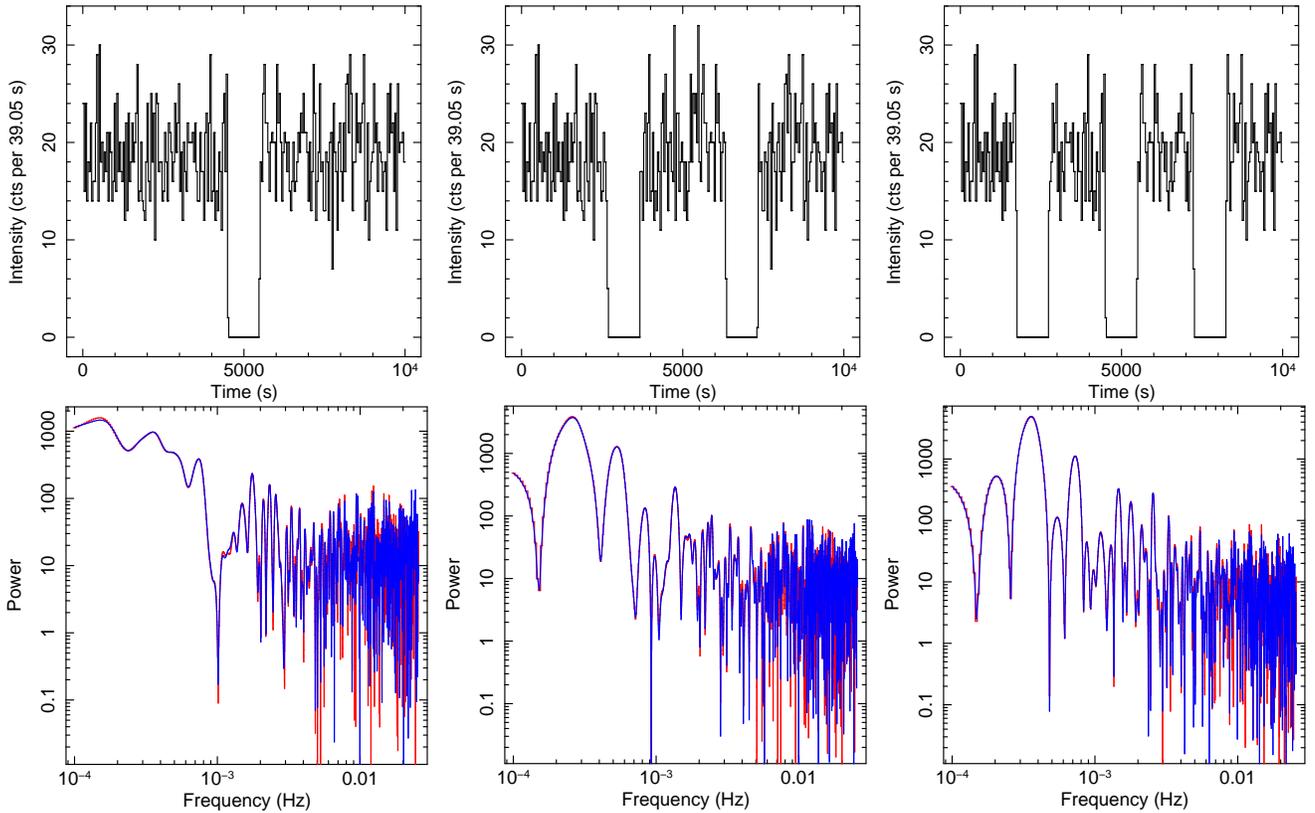

\epsscale{1.0}
\includegraphics[height=0.32 \textwidth, angle=-90]{ts_256bins_1gapsOf1000s.ps}\!\!\!%
\includegraphics[height=0.32 \textwidth, angle=-90]{ts_256bins_2gapsOf1000s.ps}\!\!\!%
\includegraphics[height=0.32 \textwidth, angle=-90]{ts_256bins_3gapsOf1000s.ps} \\
\includegraphics[height=0.32 \textwidth, angle=-90]{r2AndLS_scaled_1gapsOf1000s.ps}\!\!\!%
\includegraphics[height=0.32 \textwidth, angle=-90]{r2AndLS_scaled_2gapsOf1000s.ps}\!\!\!%
\includegraphics[height=0.32 \textwidth, angle=-90]{r2AndLS_scaled_3gapsOf1000s.ps}
\caption{\footnotesize Illustration of the effects of gaps on the periodogram of the data used in \cref{f:modR2ForEventsAndTS_withLomb}. The time series are in the top row, above their respective periodogram. The $\mathcal{R}^2_1$ statistic is in red and LS in blue. The observation duration is 10 ks, and the gaps of 1 ks in length are equally spaced.}
\label{f:gaps}
\end{figure*}

The data are those of \cref{f:modR2ForEventsAndTS_withLomb}, and we consider three cases with one, two, and three gaps equally spaced within the observation, removing a progressively larger fraction of the data, from 10\% to 30\%. \cref{f:gaps} shows the time series (top) and their periodograms (bottom; $\mathcal{R}^2_1$ in red and LS in blue) that have been normalized such that the integrated power (area under the curve) equals one, and plotted on a log--log scale to emphasize the power-law like reddening effect of the gaps. In each case the periodograms are practically indistinguishable in shape. Thus it is seen that both the $\mathcal{R}^2_1$ and the LS suffer identically from the presence of gaps in the data, and that the latter indeed does not confer any kind of protection against gap-induced distortions of the periodogram as is thought by many to be the case.

\section{Conclusion}
\label{s:conclusion}

Event data periodogram statistics are an important tool for studies of the detailed distribution in time of the detected events without having to group them, and being able to work directly with each event's recorded arrival time. This is particularly important for \xray and \gammaray pulsars. Traditional statistics commonly and currently used for such studies include the Rayleigh or $R^2$, $Z^2$, and $H$ statistics, the latter two having the former as their kernel. To maximize sensitivity to weak periodic signals, it is essential to test frequencies that are between independent Fourier frequencies. The Rayleigh, and all statistics based on it, allow for this kind of unrestricted sampling of frequency space. 

However, because of correlations in the Fourier moments for those frequencies that have a non-integer number of cycles within the observation duration, all these periodograms suffer from distorting artifacts. These distortions severely limit the sensitivity of the statistic to weak signals, especially at lower frequencies.  

More sensitive statistics must have a mechanism to take these correlations into account when computing the Fourier power. This is achieved by the modified Rayleigh statistic, $\mathcal{R}^2$, the generalization of it for an arbitrary harmonic, \rk, and the sum of more than one of the latter's components, \z.

In addition, the new \rk can be used to decompose a pulse profile by looking at the periodograms for individual harmonics. This is especially useful when some harmonics are better suited to estimate the pulse frequency and study its evolution in time, as is the case for the Crab's pulsar and most likely several other similarly fast spinning millisecond \xray pulsars. The new \z allows the possibility of summing several individual harmonics to maximize sensitivity to weak signals or to study characteristics that find their expression in certain harmonics more than in others. In the Crab pulsar, for example, most of the power is found in the second, third, and fifth harmonics. 

Having access to the information carried by each individual harmonic allows us to combine these statistically to derive the best estimate of the pulse frequency as well as an estimate of the uncertainty on this value. This is done using a weighted sum that takes into account the height and width of each periodogram's peak around the pulse frequency.

For weak, low-frequency, sinusoidal periodic signals, the more sensitive statistics are the $\mathcal{R}^2_1$ and LS that outperform the standard \fft peridogram by a factor of about two in the detection fraction at intermediate values of \sn. This is true for white and red noise. In terms of the detection fraction of false positives, the $\mathcal{R}^2_1$ and LS perform equally well and retain the same fraction independently of the sampling factor per \ifs. Gaps in the data have indistinguishable effects on both. This suggests that in the presence of data gaps, using the LS does not provide a built-in protection as is generally believed, and that careful considerations about the shape of the periodogram may require modeling tailored to the application and data.

The main conclusions of this work are that the classical $R^2$ and $Z^2$ should be replaced by \rk and \z in all applications with event data because they are far more sensitive to weak signals, and the LS should be replaced by the \rk for light curves when the uncertainties vary from one point measurement to another because their variance can have important effects on the shape of the periodogram.

\appendix

\section{The \rk Statistic: Derivation}
\label{a:modRay}

\subsection{Event Data}

The classical Rayleigh statistic can be expressed as
\begin{equation}
	R^2 = 2N(C^2 + S^2),
\label{eq:r2_appen}	  
\end{equation}
where $C$ and $S$ are defined as:
\begin{equation}
	C = \frac{1}{N} \sum_{i=1}^N \cos{\phi_i}
	\hspace{4mm} {\rm and} \hspace{4mm}
	S = \frac{1}{N} \sum_{i=1}^N \sin{\phi_i}. 
\label{eq:CandS}
\end{equation}

Assuming that $\phi$ is uniformly distributed, the probability of it having a value between 0 and 2$\pi$ is constant.  This implies that its associated {\small PDF\/}, the normalized probability density, is given by $f(\phi)$\,=\,$1/2\pi$. 

According to the central limit theorem, the sum of $N$ independent random variables $X_i$, each distributed identically about a mean $\mu$ with a finite variance $\sigma^2$, is a normal random variable distributed about a mean equal to $N\mu$ with a variance of $N\sigma^2$.  The distribution of the means $\mu_i$ also follows a normal with variance given by $\sigma^2/N$. 

In the case of the Rayleigh statistic, given that $C$ and $S$ are in fact the expectation values of $\cos{\phi_i}$ and $\sin{\phi_i}$, respectively, they are therefore distributed as normals about $\mean{\cos\phi_i}$ and $\mean{\sin\phi_i}$ with variances of $\sigma^2_C$\,=\,$Var(\cos{\phi_i})/N$ and $\sigma^2_S$\,=\,$Var(\cos{\phi})/N$. 

First, $\mean{\cos{\phi}}$\,=\,$\mean{\sin{\phi}}$\,=\,0, since the integrals of $\cos{\phi}$ and $\sin{\phi}$ from 0 to $2\pi$ are both equal to 0; therefore, $\mean{C}$\,=\,$\mean{S}$\,=\,0. Second, the variances $Var(\cos{\phi})$ and $Var(\sin{\phi})$ are equal and given by
\begin{eqnarray}
Var(\cos{\phi}) & = &  \mean{\cos^2{\phi}} - \mean{\cos{\phi}}^2 \nonumber \\
		& = &  \int_0^{2\pi} f(\phi) \cos^2{\phi} \,{\rm d}\phi  - 0 \\
		& = &  \int_0^{2\pi} \frac{1}{2\pi}\cdot\frac{1}{2}(1 + \cos{2\phi}) \,{\rm d}\phi  
			= \frac{1}{2}. \nonumber 
\end{eqnarray}
Therefore, $\displaystyle \sigma^2_C = \sigma^2_S = 1/2N$. 

Finally, since $Var(qX)$\,=\,$q^2 Var(X)$, where $q$ is a constant, the scaled variables $c$\,=\,$\sqrt{2N}\cdot\/C$ and $s$\,=\,$\sqrt{2N}\cdot\/S$, with $Var(\sqrt{2N}\cdot C) = Var(\sqrt{2N}\cdot S) = 2N\cdot \sigma^2_C = 1$, are both distributed according to the standard normal. This implies that $R^2$\,=\,$c^2 + s^2$\,=\,$2NC^2 + 2NS^2$\,=\,$2N(C^2 + S^2)$, is the sum of the squares of two normally distributed, zero-mean and unit-variance, independent variables.  The square of a standard normal variable is $\chi^2_1$ distributed, and the sum of $\chi^2$ variables is also a $\chi^2$ variable for which the number of \dof is given by the sum of the \dof of the individual variables.  Thus, $R^2$ is $\chi^2_2$ distributed. 

The frequencies within an \ifs are by definition not independent. Therefore, sampling more than one frequency per \ifs  introduces a correlation in the values of $C$ and those of $S$ within the \ifs, and thus also in the power estimates derived from them.  To remove these correlations and recover the $\chi^2_2$ distribution of powers for white noise, we must calculate the expectation values, variances, and covariance directly from the data and incorporate them in the calculation of the power. 

The modified Rayleigh statistic is defined as
\begin{equation}
\mathcal{R}^2 = 
\begin{pmatrix}
{C-\mean{C}} \\ 
{S-\mean{S}} 
\end{pmatrix}^{\rm T}
\begin{pmatrix}
\sigma^2_C & \sigma_{C S} \\
\sigma_{C S}   & \sigma^2_S
\end{pmatrix}^{-1}
\begin{pmatrix}
{C-\mean{C}} \\ 
{S-\mean{S}} 
\end{pmatrix}.
\label{eq:r2mod_appen}
\end{equation}
Note that if we replace $\mean{C}$ and $\mean{S}$ by 0, as is the case when sampling the complete phase between 0 and $2\pi$; $\sigma^2_C$ and $\sigma^2_S$ by 1, and $\sigma_{C S}$ by 0, as is the case for independent normal variables, we recover the classical Rayleigh statistic:

\begin{equation}
R^2 = 
\begin{pmatrix} c & s \end{pmatrix}
\begin{pmatrix}
 1 & 0\\
 0 & 1
\end{pmatrix}
\begin{pmatrix} c \\ s \end{pmatrix} = c^2 + s^2
\end{equation}

In constructing $\mathcal{R}^2$, we must compute \mean{C}, \mean{S}, $\sigma_C^2$, $\sigma_S^2$, and $\sigma_{C S}$.  The integral is over time and not over phase.  The duration of the observation is $T$\,=\,$t_2 - t_1$, and $\phi$ is replaced by $\omega\/t$, where $\omega$\,=\,$2\pi\nu$\,=\,$2\pi/p$, $\nu$ is the test frequency and $p$ is the corresponding test period. 

\begin{figure*}
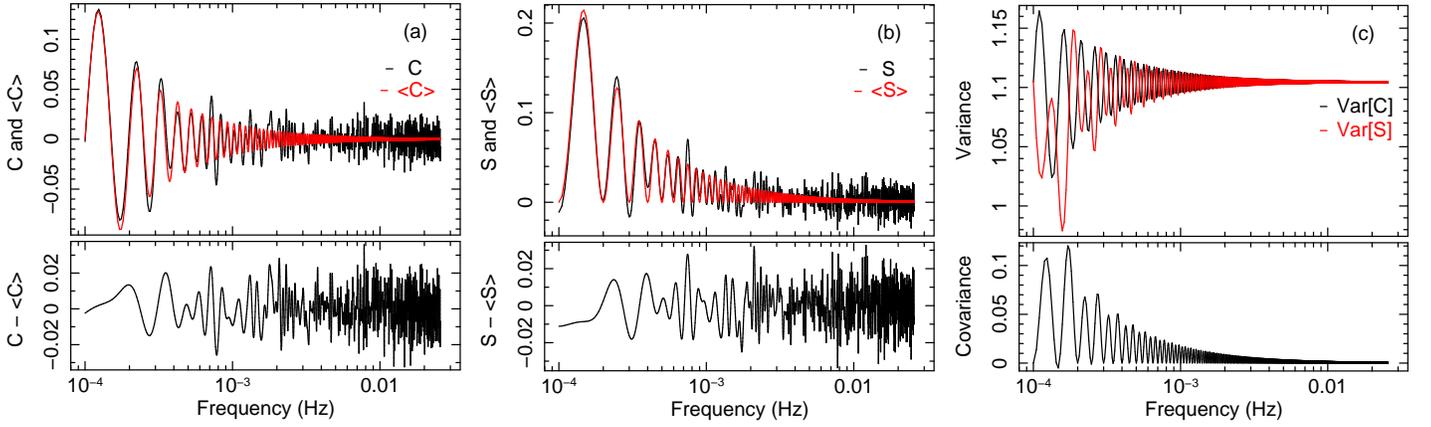

\epsscale{1.0}
\hspace{-4mm}\includegraphics[height=0.345\textwidth, angle=-90]{r2Components_cos.ps}\!\!\!%
\includegraphics[height=0.345\textwidth, angle=-90]{r2Components_sin.ps}\!\!\!%
\includegraphics[height=0.345\textwidth, angle=-90]{r2Components_var.ps}\!\!\!
\caption{\footnotesize Components of the $\mathcal{R}^2_1$ statistic for the data of \cref{f:modR2ForEventsAndTS_withLomb}. Panels (a) and (b) show, plotted as a function of frequency, the values of $C$, $\mean{C}$ and ($C-\mean{C}$), and $S$, $\mean{S}$ and ($S-\mean{S}$). Panel (c) shows the variances of $C$ and $S$ and their covariance (scaled by $2N$).}
\label{f:variables}
\end{figure*}

\begin{figure*}
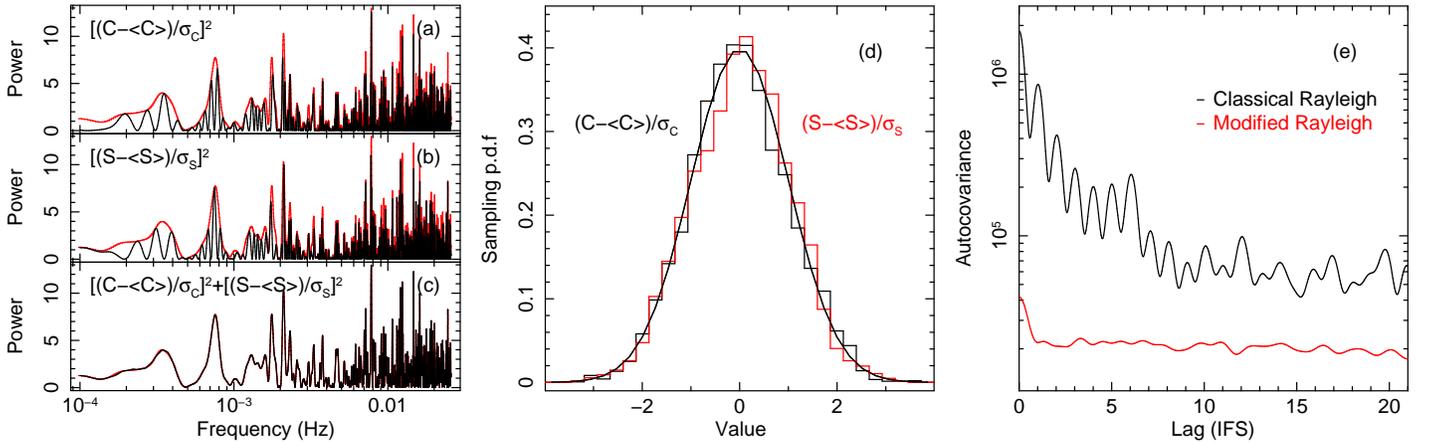

\epsscale{1.0}
\hspace{-4mm}\includegraphics[height=0.345\textwidth, angle=-90]{r2Components_terms.ps}\!\!\!%
\includegraphics[height=0.345\textwidth, angle=-90]{r2Components_histos.ps}\!\!\!%
\includegraphics[height=0.345\textwidth, angle=-90]{autocovariancesOfR2VsModR2.ps}
\caption{\footnotesize Panels (a)--(c): decomposition of the $\mathcal{R}^2_1$ into its primary components, showing the cosine term in (a), the sine term in (b), and their sum in (c). Overlaid in red is the $\mathcal{R}^2_1$ periodogram. Panel (d): histogram of the standardized variables $[(C-\mean{C})/\sigma_C]^2$ (black) and $[(S-\mean{S})/\sigma_S]^2$ (red) plotted in panels (a) and (b) with the standard normal p.d.f. (smooth black). Panel (e): autocovariances of $R^2$ (black) and $\mathcal{R}^2_1$ (red) with lag in units of independent Fourier spacings.}
\label{f:fourierComponents}
\end{figure*}

We want to generalize the expression of \cref{eq:r2mod_appen} to hold for the $k$th harmonic. Introducing the subscript $k$ on all terms to specify that they are now harmonic-specific, our matrix equation becomes
\begin{equation}
	\mathcal{R}^2_k = 
	\begin{pmatrix}
	  C_k - \mean{C_k} \\ 
	  S_k - \mean{S_k}
	\end{pmatrix}^{\rm T}
	\begin{pmatrix}
	   {\sigma^2}_{C_k} & \sigma_{C_k S_k} \\
	    \sigma_{C_k S_k}   & {\sigma^2}_{S_k}
	\end{pmatrix}^{-1}
	\begin{pmatrix}
	  C_k - \mean{C_k} \\ 
	  S_k - \mean{S_k}
	\end{pmatrix}
\label{eq:mod-r2_k_appen}
\end{equation}
In this case, the values of $C_k$ and $S_k$ defined in \cref{eq:CandS} become
\begin{equation}
	C_k = \frac{1}{N} \sum_{i=1}^N \cos{k\phi_i}
	\hspace{4mm} {\rm and} \hspace{4mm}
	S_k = \frac{1}{N} \sum_{i=1}^N \sin{k\phi_i},
\label{eq:CandS_k}
\end{equation}
with expectation values given by
\begin{eqnarray}
\mean{C_k}	& = &\frac{1}{T} \int_{t_1}^{t_2} \cos{k\omega\/t} \,{\rm d}t = \mean{\cos{k\phi} } \nonumber \\
       	& = & \frac{1}{k\omega\/T} \left( \sin{k\omega\/t_2} - \sin{k\omega\/t_1} \right), 
\end{eqnarray}
and
\begin{eqnarray}
\mean{S_k} & = & \frac{1}{T} \int_{t_1}^{t_2} \sin{k\omega\/t} \,{\rm d}t =  \mean{\sin{k\phi} }  \nonumber \\
		& = & \frac{1}{k\omega\/T} \left( \cos{k\omega\/t_1} - \cos{k\omega\/t_2} \right).
\end{eqnarray}
The variance of $C_k$ is
\begin{eqnarray}
\sigma_{C_k}^2 & = &\frac{1}{TN} \int_{t_1}^{t_2} \cos^2{k\omega\/t}\,{\rm d}t - \mean{C_k}^2 = \frac{1}{N}Var(\cos{k\phi})\nonumber \\
	   & = &\frac{1}{TN} \int_{t_1}^{t_2} \frac{1}{2} (1 + \cos{2k\omega\/t})\,{\rm d}t - \mean{C_k}^2 \nonumber \\
	   & = &\frac{1}{2TN} \left( t + \frac{\sin{2k\omega\/t}}{2k\omega} \right)_{t_1}^{t_2} - \mean{C_k}^2 \\
	    & =  &\frac{1}{2N} \left( 1+ \frac{1}{k\omega\/T} ( \sin{k\omega\/t_2}\cos{k\omega\/t_2} -  \sin{k\omega\/t_1}\cos{k\omega\/t_1} )\right) - \mean{C_k}^2. \nonumber
\end{eqnarray}
The variance of $S_k$ is
\begin{eqnarray}
\sigma_{S_k}^2  & = &\frac{1}{TN} \int_{t_1}^{t_2} \sin^2{k\omega\/t}\,{\rm d}t - \mean{S_k}^2 = \frac{1}{N}Var(\sin{k\phi}) \nonumber \\
	   &  =  &\frac{1}{TN} \int_{t_1}^{t_2} \frac{1}{2} (1 - \cos{2k\omega\/t})\,{\rm d}t - \mean{S_k}^2  \\
	   &  = &\frac{1}{2N} \left( 1- \frac{1}{k\omega\/T} ( \sin{k\omega\/t_2}\cos{k\omega\/t_2} -  \sin{k\omega\/t_1}\cos{k\omega\/t_1} ) \right) - \mean{S_k}^2. \nonumber
\end{eqnarray}
And their covariance is given by
\begin{eqnarray}
\sigma_{C_k S_k} & = & \mean{C_k\cdot S_k} - \mean{C_k} \mean{S_k} = \frac{1}{N}Cov(\cos{k\phi}, \sin{k\phi}) \nonumber \\
	& = & \frac{1}{TN} \int_{t_1}^{t_2} \cos{k\omega\/t} \sin{k\omega\/t} \,{\rm d}t - \mean{C_k} \mean{S_k} \\
	& = & \frac{1}{2k\omega\/TN} \left(\sin^2k\omega\/t_2 - \sin^2k\omega\/t_1 \right) - \mean{C_k} \mean{S_k}. \nonumber
\end{eqnarray}

Therefore, to evaluate \rk we need to compute
\begin{eqnarray}
\mean{C_k}	& = & \frac{1}{k\omega\/T} \left[ \sin{k\omega\/t} \right]_{t_1}^{t_2}, \\
\mean{S_k}	& = & \frac{-1}{k\omega\/T} \left[ \cos{k\omega\/t} \right]_{t_1}^{t_2}, \\
\sigma_{C_k}^2 & = & \frac{1}{2N} \left( 1 + \frac{1}{k\omega\/T} \left[ \sin{k\omega\/t}\cos{k\omega\/t} \right]_{t_1}^{t_2} \right) - \mean{C_k}, \\
\sigma_{S_k}^2 & = & \frac{1}{2N} \left( 1 - \frac{1}{k\omega\/T} \left[ \sin{k\omega\/t}\cos{k\omega\/t} \right]_{t_1}^{t_2} \right) - \mean{S_k}, \\
\sigma_{C_k S_k} & = & \frac{1}{2k\omega\/TN} \left[ \sin^2k\omega\/t \right]_{t_1}^{t_2} - \mean{C_k} \mean{S_k}.
\end{eqnarray}

\cref{f:variables} illustrates the importance of the effects discussed in relation to the Fourier power derived from the classical Rayleigh statistic (\cref{eq:r2_appen}) or the modified statistic (\cref{eq:r2mod_appen}, given that $k=1$). Most importantly, we see that the non-zero contribution from the analytically expected integral of sine and cosine components between independent frequencies completely dominates from mid to low frequencies, decreasing in magnitude with increasing frequency (panels (a) and (b)). In addition, note that this oscillatory behavior is also shared by the variances of $C$ and $S$ as well as by their covariance (panel (c)).

\cref{f:fourierComponents} panels (a)--(c) show, in comparison to the periodogram drawn in red, the individual contribution and sum of the squares of the standardized variables $(C-\mean{C})/\sigma_C$ and $(S-\mean{S})/\sigma_S$, making it clear that the contributions of the cosine and sine terms (panels (a) and (b)) are complementary (as expected), and that their sum accounts well for the total power in the $\mathcal{R}^2_1$ periodogram (panel (c)), which implies that the covariance, $\sigma_{CS}$, in \cref{eq:mod-r2_k_appen}, does not play a significant role here.  In panel (d) we see that subtracting the analytical expectation ($\mean{C}$ and $\mean{S}$) from the sum of Fourier moments calculated from the phases ($C$ and $S$) results in a variable distributed around zero, and that furthermore, dividing each by the expected standard deviation ($\sigma_C$ and $\sigma_S$) yields standard normal deviates, and thus effectively independent variables.

Finally, the aim of the modifications with respect to the classical Rayleigh statistic in the \rk statistic in general and in the $\mathcal{R}^2_1$ in this case is to remove the correlation in the power estimates. Autocorrelation is always maximal for a value with itself and then drops, more or less quickly depending on the extent of the correlation between neighbors, before flattening out. In the case of the classical Rayleigh periodogram, we expect the autocovariance to oscillate with a frequency matching the \ifs with an overall amplitude dropping in an exponential or power-law like fashion. In the case of the modified statistic---if the modifications successfully remove the correlation---we expect a flat curve beyond the first Fourier spacing. As shown in panel (e) of \cref{f:fourierComponents}, this is exactly what we find.

\subsection{Binned Data and Point Measurements}

We defined the \rk statistic for binned data and point measurements in \cref{eq:r2k-lc}. It was written as
\begin{equation}
	\mathcal{R}^2_k =
\frac{\left(\sum_{i=1}^{n} \rho_i \cos{k\phi_i} \right)^2}{\sum_{i=1}^{n} (\rho_i\cos{k\phi_i})^2}
+ \frac{\left(\sum_{i=1}^{n} \rho_i \sin{k\phi_i} \right)^2}{\sum_{i=1}^{n} (\rho_i \sin{k\phi_i})^2}.
\end{equation}
Expressing the weights, $\rho_i$, explicitly as $r_i/\sigma_i^2$, we get
\begin{equation}
	\mathcal{R}^2_k =
\frac{\left(\sum_{i=1}^{n} r_i/\sigma_i^2 \cos{k\phi_i} \right)^2}{\sum_{i=1}^{n} (r_i/\sigma_i^2\cos{k\phi_i})^2}
+ \frac{\left(\sum_{i=1}^{n} r_i/\sigma_i^2 \sin{k\phi_i} \right)^2}{\sum_{i=1}^{n} (r_i/\sigma_i^2 \sin{k\phi_i})^2}.
\label{eq:r2_k_explicit}
\end{equation}
If the uncertainty on the intensity measurements is equal---this is usually the case for infrared time series constructed from a set of individual image snapshots where the uncertainty on the measurements is calculated from the variance of the calibrator star's intensity over the entire set of images---then $\sigma_i$ is constant (i.e., $\sigma$). Because it is constant, this factor that appears in all the terms can be taken out of the sums and canceled out. The statistic then becomes
\begin{equation}
	\mathcal{R}^2_k =
\frac{\left(\sum_{i=1}^{n} r_i\cos{k\phi_i} \right)^2}{\sum_{i=1}^{n} (r_i\cos{k\phi_i})^2}
+ \frac{\left(\sum_{i=1}^{n} r_i\sin{k\phi_i} \right)^2}{\sum_{i=1}^{n} (r_i\sin{k\phi_i})^2}.
\end{equation}

What if the uncertainties are not equal? What is the effect of the uncertainties on the periodogram? The main point of this study was given at the end of \cref{s:modRayForLC}, namely that the size of the uncertainty on each measurement is not important in the calculation of the periodogram, but that it is the relative variations of the uncertainties, one with respect to the others, that play a role in changing the shape of the periodogram: naturally, the greater the scatter, the greater the whitening and loss of structure.

\begin{figure}[htb]
\epsscale{1.0}
\includegraphics[height=0.9\columnwidth, angle=-90]{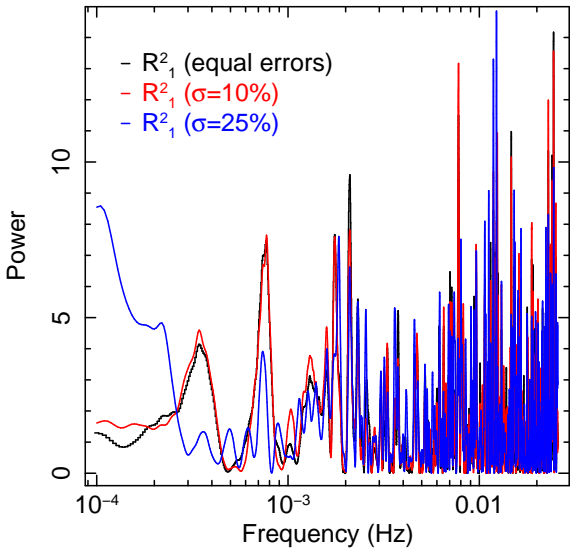}
\caption{\footnotesize Illustration of the periodogram shown in \cref{f:modR2ForEventsAndTS_withLomb} here calculated using the \rk statistic of \cref{eq:r2_k_explicit} for $k=1$, of the effects of assigning normal random uncertainties distributed with a standard deviation of 10\% (red) and 25\% (blue) to the intensity measurements. The reference in black is the unweighted periodogram.}
\label{f:uncertainties}
\end{figure}

\cref{f:uncertainties} shows the unweighted periodogram of  \cref{f:modR2ForEventsAndTS_withLomb} in black, and two additional periodograms overlaid in red and blue. In each case random normally distributed uncertainties are assigned to the individual intensity measurements in each bin. The first shows the effect of a standard deviation of 10\% in the distribution of uncertainties, and we see that the deviations from the black periodogram are small. The second shows the effects of a 25\% standard deviation on the uncertainties where differences are noticeably more important.

Such a periodogram statistic in which each measurement is weighted by a potentially different uncertainty is essential in applications where we are interested in combining data from different observations, different instruments, or both, because the uncertainties from one observation to another and from one instrument to another will inevitably vary.

It is also essential when treating time series that even if from a single instrument and a single continuous observation may vary in quality from bin to bin, as is the case for the \gammaray observatory \emph{INTEGRAL}, because each estimate of intensity in each bin is derived from independent snapshots whose statistical properties vary depending on factors such as the duration of the snapshot, the pointing's direction in the sky, the position within the satellite in its orbit, the time from the passage through the radiation belts, etc. In such cases, as well as in similar ones, it would clearly be a mistake to rely on the unweighted periodogram when so many factors contribute differently to the quality of each data point in the time series.

\bibliography{refs-final}

\begin{thebibliography}{}
\providecommand\natexlab[1]{#1}
\providecommand\JournalTitle[1]{#1}

\bibitem[{{Belloni} \& {Hasinger}(1990)}]{1990A&A...227L..33B}
{Belloni}, T., \& {Hasinger}, G. 1990, \JournalTitle{\aap},
  \href{http://adsabs.harvard.edu/abs/1990A\%26A...227L..33B}{227},
  \href{http://adsabs.harvard.edu/abs/1990A\%26A...227L..33B}{L33}

\bibitem[{{Bozzo} {et~al.}(2012){Bozzo}, {Ferrigno}, {T{\"u}rler},
  {Manousakis}, \& {Falanga}}]{2012A&A...545A..83B}
{Bozzo}, E., {Ferrigno}, C., {T{\"u}rler}, M., {Manousakis}, A., \& {Falanga},
  M. 2012,
  \href{http://dx.doi.org/10.1051/0004-6361/201219344}{\JournalTitle{\aap},
  \href{http://adsabs.harvard.edu/abs/2012A\%26A...545A..83B}{545},
  \href{http://adsabs.harvard.edu/abs/2012A\%26A...545A..83B}{A83}}

\bibitem[{{Buccheri} {et~al.}(1983){Buccheri}, {Bennett}, {Bignami}, {Bloemen},
  {Boriakoff}, {Caraveo}, {Hermsen}, {Kanbach}, {Manchester}, {Masnou},
  {Mayer-Hasselwander}, {Ozel}, {Paul}, {Sacco}, {Scarsi}, \&
  {Strong}}]{1983A&A...128..245B}
{Buccheri}, R., {Bennett}, K., {Bignami}, G.~F., {et~al.} 1983,
  \JournalTitle{\aap},
  \href{http://adsabs.harvard.edu/abs/1983A\%26A...128..245B}{128},
  \href{http://adsabs.harvard.edu/abs/1983A\%26A...128..245B}{245}

\bibitem[{{Burderi} {et~al.}(2006){Burderi}, {Di Salvo}, {Menna}, {Riggio}, \&
  {Papitto}}]{2006ApJ...653L.133B}
{Burderi}, L., {Di Salvo}, T., {Menna}, M.~T., {Riggio}, A., \& {Papitto}, A.
  2006, \href{http://dx.doi.org/10.1086/510666}{\JournalTitle{ApJL},
  \href{http://adsabs.harvard.edu/abs/2006ApJ...653L.133B}{653},
  \href{http://adsabs.harvard.edu/abs/2006ApJ...653L.133B}{L133}}

\bibitem[{{de Jager}(1994)}]{1994ApJ...436..239D}
{de Jager}, O.~C. 1994,
  \href{http://dx.doi.org/10.1086/174896}{\JournalTitle{ApJ},
  \href{http://adsabs.harvard.edu/abs/1994ApJ...436..239D}{436},
  \href{http://adsabs.harvard.edu/abs/1994ApJ...436..239D}{239}}

\bibitem[{{de Jager} {et~al.}(1986){de Jager}, {Raubenheimer}, \&
  {Swanepoel}}]{1986A&A...170..187D}
{de Jager}, O.~C., {Raubenheimer}, B.~C., \& {Swanepoel}, J.~W.~H. 1986,
  \JournalTitle{\aap},
  \href{http://adsabs.harvard.edu/abs/1986A\%26A...170..187D}{170},
  \href{http://adsabs.harvard.edu/abs/1986A\%26A...170..187D}{187}

\bibitem[{{de Jager} {et~al.}(1989){de Jager}, {Raubenheimer}, \&
  {Swanepoel}}]{1989A&A...221..180D}
{de Jager}, O.~C., {Raubenheimer}, B.~C., \& {Swanepoel}, J.~W.~H. 1989,
  \JournalTitle{\aap},
  \href{http://adsabs.harvard.edu/abs/1989A\%26A...221..180D}{221},
  \href{http://adsabs.harvard.edu/abs/1989A\%26A...221..180D}{180}

\bibitem[{{de Rosa} {et~al.}(2009){de Rosa}, {Ubertini}, {Campana}, {Mazzano},
  {Dean}, \& {Bassani}}]{2009MNRAS.393..527D}
{de Rosa}, A., {Ubertini}, P., {Campana}, R., {et~al.} 2009,
  \href{http://dx.doi.org/10.1111/j.1365-2966.2008.14160.x}{\JournalTitle{\mnras},
  \href{http://adsabs.harvard.edu/abs/2009MNRAS.393..527D}{393},
  \href{http://adsabs.harvard.edu/abs/2009MNRAS.393..527D}{527}}

\bibitem[{{Leahy} {et~al.}(1983{\natexlab{a}}){Leahy}, {Darbro}, {Elsner},
  {Weisskopf}, {Kahn}, {Sutherland}, \& {Grindlay}}]{1983ApJ...266..160L}
{Leahy}, D.~A., {Darbro}, W., {Elsner}, R.~F., {et~al.} 1983{\natexlab{a}},
  \JournalTitle{ApJ},
  \href{http://adsabs.harvard.edu/abs/1983ApJ...266..160L}{266},
  \href{http://adsabs.harvard.edu/abs/1983ApJ...266..160L}{160}

\bibitem[{{Leahy} {et~al.}(1983{\natexlab{b}}){Leahy}, {Elsner}, \&
  {Weisskopf}}]{1983ApJ...272..256L}
{Leahy}, D.~A., {Elsner}, R.~F., \& {Weisskopf}, M.~C. 1983{\natexlab{b}},
  \href{http://dx.doi.org/10.1086/161288}{\JournalTitle{ApJ},
  \href{http://adsabs.harvard.edu/abs/1983ApJ...272..256L}{272},
  \href{http://adsabs.harvard.edu/abs/1983ApJ...272..256L}{256}}

\bibitem[{{Meyer} {et~al.}(2008){Meyer}, {Do}, {Ghez}, {Morris}, {Witzel},
  {Eckart}, {B{\'e}langer}, \& {Sch{\"o}del}}]{2008ApJ...688L..17M}
{Meyer}, L., {Do}, T., {Ghez}, A., {et~al.} 2008,
  \href{http://dx.doi.org/10.1086/593147}{\JournalTitle{ApJL},
  \href{http://adsabs.harvard.edu/abs/2008ApJ...688L..17M}{688},
  \href{http://adsabs.harvard.edu/abs/2008ApJ...688L..17M}{L17}}

\bibitem[{{Miyamoto} {et~al.}(1991){Miyamoto}, {Kimura}, {Kitamoto}, {Dotani},
  \& {Ebisawa}}]{1991ApJ...383..784M}
{Miyamoto}, S., {Kimura}, K., {Kitamoto}, S., {Dotani}, T., \& {Ebisawa}, K.
  1991, \href{http://dx.doi.org/10.1086/170837}{\JournalTitle{ApJ},
  \href{http://adsabs.harvard.edu/abs/1991ApJ...383..784M}{383},
  \href{http://adsabs.harvard.edu/abs/1991ApJ...383..784M}{784}}

\bibitem[{{Orford}(1996)}]{1996APh.....4..235O}
{Orford}, K.~J. 1996, \JournalTitle{Astroparticle Physics},
  \href{http://adsabs.harvard.edu/abs/1996APh.....4..235O}{4},
  \href{http://adsabs.harvard.edu/abs/1996APh.....4..235O}{235}

\bibitem[{{Press} {et~al.}(2002){Press}, {Teukolsky}, {Vetterling}, \&
  {Flannery}}]{2002nrc..book.....P}
{Press}, W.~H., {Teukolsky}, S.~A., {Vetterling}, \& {Flannery}, B.~P. 2002,
  \href{http://adsabs.harvard.edu/abs/2002nrc..book.....P}{Numerical Recipes in
  C++ : the Art of Scientific Computing} (Cambridge: Cambridge Univ. Press)

\bibitem[{{Romano} {et~al.}(2010){Romano}, {Sidoli}, {Ducci}, {Cusumano}, {La
  Parola}, {Pagani}, {Page}, {Kennea}, {Burrows}, {Gehrels}, {Sguera}, \&
  {Mazzano}}]{2010MNRAS.401.1564R}
{Romano}, P., {Sidoli}, L., {Ducci}, L., {et~al.} 2010,
  \href{http://dx.doi.org/10.1111/j.1365-2966.2009.15789.x}{\JournalTitle{\mnras},
  \href{http://adsabs.harvard.edu/abs/2010MNRAS.401.1564R}{401},
  \href{http://adsabs.harvard.edu/abs/2010MNRAS.401.1564R}{1564}}

\bibitem[{Royall(1997)}]{Royall:1997vc}
Royall, R.~M. 1997,
  \href{http://www.crcpress.com/product/isbn/9780412044113}{Statistical
  Evidence, A Likelihood Paradigm} (New York: Chapman {\&} Hall/CRC)

\bibitem[{{Scargle}(1982)}]{1982ApJ...263..835S}
{Scargle}, J.~D. 1982,
  \href{http://dx.doi.org/10.1086/160554}{\JournalTitle{ApJ},
  \href{http://adsabs.harvard.edu/abs/1982ApJ...263..835S}{263},
  \href{http://adsabs.harvard.edu/abs/1982ApJ...263..835S}{835}}

\bibitem[{{Scargle}(1989)}]{1989ApJ...343..874S}
{Scargle}, J.~D. 1989,
  \href{http://dx.doi.org/10.1086/167757}{\JournalTitle{ApJ},
  \href{http://adsabs.harvard.edu/abs/1989ApJ...343..874S}{343},
  \href{http://adsabs.harvard.edu/abs/1989ApJ...343..874S}{874}}

\bibitem[{{Vio} {et~al.}(2013){Vio}, {Diaz-Trigo}, \&
  {Andreani}}]{2013A&C.....1....5V}
{Vio}, R., {Diaz-Trigo}, M., \& {Andreani}, P. 2013,
  \href{http://dx.doi.org/10.1016/j.ascom.2012.12.001}{\JournalTitle{A\&C},
  \href{http://adsabs.harvard.edu/abs/2013A&C.....1....5V}{1},
  \href{http://adsabs.harvard.edu/abs/2013A&C.....1....5V}{5}}

\end{thebibliography}

\end{document}